\documentclass[a4paper,twocolumn,11pt,]{quantumarticle} \pdfoutput=1
\usepackage[utf8]{inputenc}
\usepackage[english]{babel}
\usepackage[T1]{fontenc}
\usepackage{amsmath}
\usepackage{hyperref}
\usepackage{wrapfig}

\usepackage{tikz}
\usepackage{lipsum}
\usepackage{subfigure}

\usepackage{hyperref}
\newcommand{\unit}[1]{\ensuremath{\, \mathrm{#1}}}

\usepackage{tikz}
\usepackage{lipsum}
\definecolor{ao(english)}{rgb}{0.0, 0.5, 0.0}
\usetikzlibrary{calc,angles,quotes}
\usepackage{enumerate}
\usepackage{amssymb}
\usepackage{epstopdf}

\usepackage{xcolor}
\usepackage{mathdots}
\usepackage{mathrsfs}
\usepackage{amsthm}
\usepackage{dcolumn}
\usepackage{bm}
\newtheorem{theorem}{Theorem}[section]
\newtheorem{lemma}[theorem]{Lemma}
\newtheorem{proposition}[theorem]{Proposition}

\theoremstyle{definition}
\newtheorem{definition}[theorem]{Definition}

\newtheorem{remark}[theorem]{Remark}
\newcommand{\mcb}{\mathscr{B}}
\newcommand{\mch}{\mathcal{H}}

\usepackage{tikz}
\usetikzlibrary{calc,angles,quotes}
\usepackage{enumerate}
\usepackage{amssymb}
\usepackage{epstopdf}

\usepackage{xcolor}
\usepackage{mathdots}
\usepackage{amsthm}
\usepackage{dcolumn}
\usepackage{bm}
\usepackage{tabularx}
\usepackage{qcircuit}
\usepackage{subfigure}

\theoremstyle{definition}

\begin{document}
\title{Reverse Map Projections as Equivariant Quantum \newline Embeddings}

\author{Max Arnott}
\affiliation{Zaiku Group Ltd., 7th Floor, 4 St Paul's Square, Liverpool L3 9SJ, UK}

\author{Dimitri Papaioannou}
\affiliation{Tecacet Inc., San Francisco, CA, USA}

\author{Kieran McDowall}

\affiliation{National Quantum Computing Centre, Rutherford Appleton Laboratory, Didcot OX11 0FA, UK
}

\author{Phalgun Lolur}

\affiliation{National Quantum Computing Centre, Rutherford Appleton Laboratory, Didcot OX11 0FA, UK
}

\author{Bambord\'{e} Bald\'{e}}
\affiliation{Zaiku Group Ltd., 7th Floor, 4 St Paul's Square, Liverpool L3 9SJ, UK}

\maketitle

\begin{abstract}
We introduce the novel class $(E_\alpha)_{\alpha \in [-\infty,1)}$ of \hbox{\textit{reverse map projection}} embeddings, each one defining a unique new method of encoding classical data into quantum states. Inspired by well-known map projections from the unit sphere onto its tangent planes, used in practice in cartography, these embeddings address the common drawback of the amplitude embedding method, wherein scalar multiples of data points are identified and information about the norm of data is lost.

We show how reverse map projections can be utilised as equivariant embeddings for quantum machine learning. Using these methods, we can leverage symmetries in classical datasets to significantly strengthen performance on quantum machine learning tasks. 

Finally, we select four values of $\alpha$ with which to perform a simple classification task, taking $E_\alpha$ as the embedding and experimenting with both equivariant and non-equivariant setups. We compare their results alongside those of standard amplitude embedding.
\end{abstract}

\section{Introduction}

%-----------------------------------------------------------

In today's noisy, intermediate scale quantum (NISQ) era of quantum computing, quantum algorithms face challenges from hardware limitations, including qubit quantity, connectivity, and noise \cite{preskill}. To reduce the negative effects of these challenges for the purpose of quantum machine learning (QML), they must be considered at every step of the QML process: data embedding, computation, and measurement. Particularly, clever choices of `classical-to-quantum' embedding schemes and efficient quantum circuits are essential to maximise the utility of current quantum resources.

The mapping of a classical dataset $\mathcal{D}$ to a set of quantum states is the necessary first step for said data to be processed by a quantum computer. Precisely, we are required to construct some map $E$ from $\mathcal{D}$ to the unit sphere of a Hilbert space $\mch$, which is the state space of the given quantum computer. Popular embedding schemes in quantum information science include basis encoding, amplitude embedding, and angle embedding. Basis encoding works by taking a datapoint $b$ expressed in binary, and assigning it the quantum state $|b\rangle$, which belongs to the standard orthonormal basis of the quantum computer. Amplitude embedding maps a classical vector $x=(x_i)_{i=0}^{2^n-1}$ to the quantum state \[\|x\|_2^{-1}\cdot\sum_{i=0}^{2^n-1} x_i |i\rangle \;.\] Angle embedding uses rotation gates to encode classical data into a quantum state, by mapping a real vector $(x_i)_{i=0}^n$ to the state \[ \bigotimes_{i=0}^n R_X(x_i) |0\rangle\]   \cite{NC,khan2024beyond, rath}.

A projection on a Hilbert space $\mch$ is always surjective when restricted to its complemented range. As such, embeddings can often be thought of as partial inverses to projections. Predefined projections on Hilbert spaces, and their domain restrictions to the unit sphere, can therefore serve naturally as inspiration for those looking to define new quantum embedding schemes. 

Cartographers have defined a plethora of projections to send the three dimensional surface of the earth onto a two dimensional map (the word `map' now being used here in the cartographic sense) \cite{snyder, MapProj}. Due to inherent complexities in translating a sphere's surface to a flat map, distortions of geometric properties are unavoidable. However, many of these projections are tailored to preserve or partially preserve certain features of the sphere to suit particular mapping requirements. Further, many of these projections can also be easily generalised to higher dimensions. A key objective of this paper is to use the mathematical definitions of such projections to construct a class of classical-to-quantum data embeddings, and apply them in practice to a data classification task.

As previously mentioned, when constructing efficient QML algorithms for the NISQ era, particular attention also needs to be paid to the structure of the variational quantum circuit. Effective circuit design is going to largely depend on the nature of the data at hand. \textit{Geometric quantum machine learning} is an emerging area of study which aims to utilise the geometry of data to inform circuit design \cite{perrier}. For example, if the data in question has a particular symmetry, then it is possible to construct a quantum circuit for which symmetrically related input states always yield symmetrically related outputs. This is the essence of the \textit{equivariant embeddings} method. See e.g., \cite{meyer, nguyen, west, schatzki} for examples of this method in action, and demonstrations of how it can improve QML performance. Once our cartography-inspired quantum embeddings are defined, we show that they can always be used as equivariant embeddings for equivariant circuits, for datasets with any finite symmetry group. 

This paper is structured as follows. In Section \ref{sectionBackground} we discuss the mathematical fundamentals required for our constructions. Alongside the basics of Hilbert spaces and their bounded operators, unitary representations of finite groups are required.

Section \ref{sectionMapProjection} details some well-known projections that are used to map spheres to planes, and generalises them to higher dimensions. The key result of the section is Proposition \ref{propDomainRange}, which defines suitable domain and codomain restrictions to make such that these projections become bijections from a subset of the unit hypersphere.

Section \ref{sectionEmbeddingDefn} focuses on the inverses of the aforementioned bijective restricted projections, which we term \textit{reverse map projections}. These functions are augmented such that their domain is the entirety of $\mathbb{R}^n$. The section concludes by describing how reverse map projections can be used as data embeddings for quantum computing.

The objective of Section \ref{sectionEquivariance} is to prove Theorem \ref{EalphaIsEquivariant}, which tells us that reverse map projections are equivariant with with respect to any given unitary group representation. Along the way, we explain how a group representation can be used to find group equivariant gates with which to construct an equivariant quantum circuit. 

Finally, this paper concludes with Section \ref{sectionExperiments}, where we compare four different reverse map projections to one another, and to amplitude embedding, by analysing their performances on a basic classification task. We use both equivariant and non-equivariant models to make our comparisons.

\section{Preliminaries}\label{sectionBackground}
Here, we explain in detail the mathematical terminology, conventions, and notation used throughout. We remark that the notation $\langle\, \cdot\,,\,\cdot\,\rangle$ has been chosen to denote inner products in this paper as opposed to the Dirac notation $\langle \,\cdot \,| \,\cdot\, \rangle$ more commonly seen in quantum-related material. Due to the pure mathematical nature of large portions of our research, we decided to align our notation to the pure mathematical convention.

\subsection{Basics}
We write $\mathbb{N}$ for the set of \textit{natural numbers}, $\mathbb{R}$ for the set of \textit{real numbers}, and $\mathbb{C}$ for the set of \textit{complex numbers}. Henceforth, the notation $\mathbb{K}$ may denote either field $\mathbb{R}$ or $\mathbb{C}$. 

Let $V$ be a vector space over $\mathbb{K}$, with norm $\|\cdot\|$. The \textit{closed unit ball} of $V$ is the set \[B_V = \{ v \in V : \|v\| \leq 1\}\;.\] The \textit{unit sphere} $\mathbb{S}_V$ of $V$ is the set of exterior points of $B_V$, that is, \[\mathbb{S}_V = \{v \in V : \|v\| = 1\} \;.\]

A \textit{Hilbert space} $\mathcal{H}$ is an inner product space which is complete with respect to the norm derived from its inner product. Completeness in this context is the property that every Cauchy sequence in $\mathcal{H}$ is convergent, with limit in $\mathcal{H}$. For every $n \in \mathbb{N}$, the space $\mathbb{K}^n$ is a Hilbert space with respect to the inner product \[ \langle (x_i)_{i=0}^{n-1}, (y_i)_{i=0}^{n-1} \rangle = \sum_{i=0}^{n-1} x_i \overline{y_i}      \] for every $(x_i)_{i=0}^{n-1}, (y_i)_{i=0}^{n-1} \in \mathbb{K}^n$, which yields the norm \[  \|(x_i)_{i=0}^{n-1}\|_2 = \left(\sum_{i=0}^{n-1} |x_i|^2\right)^{\frac{1}{2}}\;.     \] 

From here on, we drop the subscript $2$ when using the above norm. Throughout this paper, when $n \in \mathbb{N}$ is given, we use the convention that vectors in $n+1$ dimensional spaces are indexed by the set $\{0,1,\dots,n\}$ whereas vectors in $n$ dimensional spaces are indexed by $\{0,1,\dots,n-1\}$.

\subsection{Operators}\label{subsectionOperators}
A \textit{bounded operator} from a Hilbert space $\mathcal{H}$ to a Hilbert space $\mathcal{G}$ is a continuous, linear function $\mch \to \mathcal{G}$. We write $\mcb(\mch;\mathcal{G})$ for the vector space of all bounded operators from $\mch$ to $\mathcal{G}$, and we abbreviate $\mcb(\mch;\mch)$ to $\mcb(\mch)$.

Let $T \in \mcb(\mch; \mathcal{G})$ for some Hilbert spaces $\mch$ and $\mathcal{G}$. We write $T^*\in \mcb(\mathcal{G};\mathcal{H})$ to denote the \textit{adjoint} (or \textit{dual}) of $T$, that is, the unique operator for which \[\langle Tx,y\rangle = \langle x, T^*y\rangle \;,\] for every $x \in \mch$ and every $y \in \mathcal{G}$.

An operator $U\in \mcb(\mch)$ on a Hilbert space $\mch$ is \textit{unitary} if \[UU^* = U^*U = I_\mathcal{H}\;,\] where $I_\mathcal{H}$ is the \textit{identity} operator on $\mch$ with action $x\mapsto x$ for every $x \in \mch$. Unitary operators $U$ on $\mch$ are \textit{isometries}, meaning that for every $x \in \mch$, we have $\|Ux\| =\|x\|$.

A \textit{projection} on a Hilbert space $\mathcal{H}$ is an operator $P\in \mcb(\mch)$ for which \[P^2=P^*=P\;.\] The range $P(\mathcal{H})$ of $P$ is then a complemented subspace of $\mathcal{H}$, and thus, when convenient, we may identify a projection $P$ with its corestriction to $P(\mathcal{H})$, so that $P \in \mcb(\mch;P(\mch))$. We shall always make clear the domain and codomain of all projections, so this will not cause confusion.

\subsection{Group Representations}
Let $G$ be a nonempty set, equipped with a binary operation $\,\cdot\,$. We say that $G$ is a \textit{group} under $\,\cdot\,$ if it satisfies the following three conditions.
    \begin{itemize}\item \textit{Associativity:}     \[ \forall f, g, h \in G, \quad (f \cdot g) \cdot h = f \cdot (g \cdot h) \;.\]

    \item \textit{Identity Element:}
    \[ \exists e \in G \, \text{ such that } \, \forall g \in G, \quad e \cdot g = g \cdot e = g \;.\]
    
    \item \textit{Inverse Elements:} 
    \begin{equation*}\begin{gathered}
        \forall g \in G, \, \exists g^{-1} \in G \, \text{ such that } \\g \cdot g^{-1} = g^{-1} \cdot g = e\;. 
    \end{gathered}
    \end{equation*} \end{itemize}

Let $G$ be a group, and let $\mathcal{H}$ be a Hilbert space. A \textit{representation} of $G$ on $\mathcal{H}$ is a function $\rho : G \to \mcb(\mch)$ for which $\rho(g)$ is invertible (i.e., bijective) and \[\rho(g)\rho(h) = \rho(g\cdot h)\] for every $g,h \in G$. If we further have that $\rho(g)$ is a unitary operator for every $g\in G$, we say that $\rho$ is a \textit{unitary representation}. For a unitary representation $\rho$ of $G$, we also have that \[ \rho(g)^* = \rho(g^{-1})\] for every $g \in G$. If $\rho(g) = I_\mathcal{H}$ for every $g \in G$, then $\rho$ is called the \textit{trivial representation} of $G$ on $\mathcal{H}$.

Let $\rho_1$ and $\rho_2$ be representations of a group $G$ on two Hilbert spaces $\mathcal{H}_1$ and $\mathcal{H}_2$ respectively. The \textit{direct sum} $\rho_1 \oplus \rho_2$ of $\rho_1$ and $\rho_2$ is the representation of $G$ on $\mch_1 \oplus \mch_2$ with action \[ (\rho_1 \oplus \rho_2)(g) (x_1 \oplus x_2) =    \rho_1(g) x_1 \oplus \rho_2(g) x_2 \] for every $g \in G, x_1 \in \mch_1$, and $x_2 \in \mch_2$.

See \cite{ragone} for more information on representation theory in the context of QML.

\section{Higher Dimensional Map Projections}\label{sectionMapProjection}

A \textit{map projection}, loosely defined, is any projection on $\mathbb{R}^3$ which is used primarily to map some subset of a sphere to a subset of a plane. Map projections are named for their use in cartography for creating maps of the earth \cite{snyder}. For myriad topological reasons, distortions of certain geometric properties of a sphere are inevitable under a map projection. A wide array of different map projections have been defined, specifically designed in order to best preserve certain features of the sphere, for different mapping needs. Oftentimes this comes with the trade-off of increasing the distortion of other aspects.

Map projections $P_p$ can be defined by taking some \textit{centre of projection} $p$, and acting on each point $x$ of the sphere by mapping $x$ to the unique point $P_px$ on a given tangent plane of the sphere for which $P_px$ is colinear with both $x$ and $p$. In this paper, we highlight four particular map projections. Below, we give a brief overview of each one. In each case, the maps are $\mathbb{R}^3 \to \mathbb{R}^3$, the given sphere is $\mathbb{S}_{\mathbb{R}^3}$, and the tangent plane in question is the set $\{(x_0,x_1,x_2)\in \mathbb{R}^3 : x_2 = 1\}$.

\begin{itemize}
    \item \textit{Gnomonic:} This projection sets the centre of projection as the origin. Under a gnomonic projection, great circles are mapped to straight lines, a feature useful in e.g., aviation. A drawback of the gnomonic projection is that it causes high distortion to shape and area, especially towards the edge of the map. 
    
    \item \textit{Stereographic:} The centre of projection for the stereographic projection is the point $(0,0,-1)$, which is antipodal to the point of contact between the sphere and the tangent plane. Locally, angles and shapes are well-preserved under this projection. This preservation of angles is a helpful feature in navigation.
    
    \item \textit{Twilight:} The twilight projection places the centre of projection outside of the sphere, at the point $(0,0,-1-\frac{\sqrt{2}}{2})$. This projection provides somewhat of a compromise between the stereographic and orthographic views in terms of its preservation of geometric features.
    
    \item \textit{Orthographic:} Finally, the orthographic projection sets the point of projection at $(0,0,-\infty)$ (mathematically speaking, the operator $P_{-\infty}$ is the limit in operator norm of the sequence of operators $(P_{n})$ as $n \to -\infty$). This map projection is \textit{azimuthal}, which means that directions from the centre point to any other point on the map are preserved. There is however, high distortion to shape and area. \end{itemize}

For a detailed exploration of map projections, see \cite{MapProj}.

In this paper, we generalise the term `map projection' to higher dimensions, enabling the definition to include projections of hyperspheres of arbitrary finite dimension to their tangent hyperplanes. We shall now introduce the notation required to handle such projections. For a subset $W$ of a vector space $V$, and some element $p\in V$, we write $W-p$ to denote the subset $\{w-p : w\in W\}$ of $V$. 

\begin{definition}\label{defnPWp}
 Let $\mathcal{H}$ be a Hilbert space over a field $\mathbb{K}$. Fix some constants $a \in \mathcal{H}$ and $b \in \mathbb{K}$, and consider the hyperplane \[W = \{ x \in \mch : \langle a,x\rangle = b \}\] of $\mch$. 

 The \textit{projection from $0$ to $W$} is defined as the operator $P_0^W : \mch \to \mch$ with action
 \[x \mapsto x - \frac{\langle a,x\rangle - b}{\|a\|^2}a \] for all $x \in \mch$.

 Fix a point $p \in \mch$. The \textit{projection from $p$ to $W$} is then defined as the function $P_p^W : \mch \to \mch$ with action \[ x \mapsto P_0^{W-p} x + p\] for every $x \in \mch$. 
\end{definition}

\begin{remark}
    Definition \ref{defnPWp} does not typically define a projection in the operator-theoretic sense since its span is an affine subspace of $\mathcal{H}$ rather than a true subspace whenever $W$ does not contain the origin. This can prevent the function $P_p^W$ from being linear. 
\end{remark}

From here on, we will consider only map projections from a point of projection along the $X_n$ axis of $\mathbb{R}^{n+1}$ onto the tangent hyperplane $W$ defined by equation $x_{n}=1$. It will be helpful for us to identify the hyperplane $W$ with $\mathbb{R}^n$, which will be done by ignoring the final coordinate of each point in $W$, which is always equal to $1$. This process is incorporated into Definition \ref{PalphaDefinition} below, which provides us with a class \[\{P_\alpha : \alpha \in [-\infty,1) \}\] of map projections, containing the four map projections discussed earlier.

\begin{definition}\label{PalphaDefinition} Let $\alpha \in (-\infty , 1)$ and let $n \in \mathbb{N}$. Let \[W = \{ (x_i)_{i=0}^{n} \in \mathbb{R}^{n+1} : x_{n} = 1\}  \;,\] and set $p = (0,0,\dots,0,\alpha) \in \mathbb{R}^{n+1}$. Then define $P_\alpha \in \mcb(\mathbb{R}^{n+1};\mathbb{R}^n)$ to have action \[ P_\alpha x = (P_p^W x)_{i=0}^{n-1}\] for every $x \in \mathbb{R}^{n+1}$, where $P_p^W$ is in the notation of Definition \ref{defnPWp}.

Further, define $P_{-\infty} \in \mcb(\mathbb{R}^{n+1};\mathbb{R}^n)$ to have action \[(x_i)_{i=0}^n \mapsto (x_i)_{i=0}^{n-1}\] for every $(x_i)_{i=0}^n\in \mathbb{R}^{n+1}$.
\end{definition}

The gnomonic, stereographic, twilight, and orthographic map projections can all be expressed in terms of Definition \ref{PalphaDefinition}. Their definitions all take $n=2$ and their values of $\alpha$ are $0, -1, (-1-\frac{\sqrt{2}}{2})$, and $-\infty$ respectively. To help us in Section \ref{sectionEmbeddingDefn} to define the embeddings $E_\alpha$ we desire, we give Lemma \ref{concreteDefnPalpha} and Proposition \ref{propDomainRange} below.

Lemma \ref{concreteDefnPalpha} reveals to us the concrete coordinates of the images of points on the unit sphere $\mathbb{S}_{\mathbb{R}^{n+1}}$ under $P_\alpha$, which will aid us in future calculations, and potentially make the definition of $P_\alpha$ more intuitive to the reader by removing abstraction. Using that information, we prove Proposition \ref{propDomainRange}, which tells us domain and codomain restrictions of $P_\alpha$ to make in order to obtain a bijection. This is of course vital information when one wishes to obtain a partial inverse of $P_\alpha$, which is our main objective for Section \ref{sectionEmbeddingDefn}.

\begin{lemma}\label{concreteDefnPalpha} Let $x=(x_0,x_1,\dots,x_n)\in \mathbb{S}_{\mathbb{R}^{n+1}}$ for some $n \in \mathbb{N}$, let $\alpha \in (-\infty,1)$ and let $P_\alpha \in \mcb(\mathbb{R}^{n+1};\mathbb{R}^n)$ be as in Definition \ref{PalphaDefinition}. Then \[ P_\alpha x = \left(\frac{1-\alpha}{x_n-\alpha}\right)(x_0,x_1,\dots,x_{n-1})\;.\]
\begin{proof}

Set $P_\alpha x = (x_i')_{i=0}^{n-1}$, and set $i \in \{0,\dots, n-1\}$. Recall that $P_\alpha x $ can be defined as the unique point in $\mathbb{R}^{n+1}$ with $n$th coordinate equal to $1$, colinear with both $x$ and $(0 , \dots, 0, \alpha)$, with its $n$th coordinate then removed. 
 When viewing $P_\alpha x$ through this lens, as pictured in Figure \ref{fig:trianglepic}, we see that \begin{equation} \label{triangleeqn}
     \frac{x_i'}{1-\alpha} = \frac{x_i}{x_n - \alpha}\;. 
 \end{equation} The result follows. \qedhere
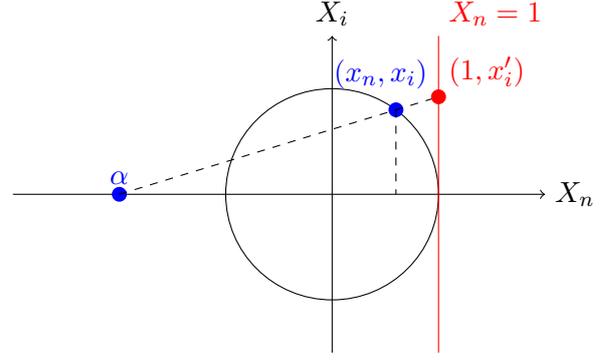
\begin{figure}
    \centering
    \begin{tikzpicture}[scale=1.4]
    % Draw the axes
    \draw[->] (-3,0) -- (2,0) node[right] {$X_n$};
    \draw[->] (0,-1.5) -- (0,1.5) node[above] {$X_i$};
    
    % Draw the circle of radius 1 centered at the origin
    \draw (0,0) circle (1);
    
    % Define the point on the circle (example point (0.6,0.8) for illustration)
    \coordinate (A) at (0.6, 0.8);
    
    % Draw the point on the circle and label it
    \fill[blue] (A) circle (2pt) node[below, shift={(-0.2,+0.8)}] {$(x_n, x_i)$};
    
    % Draw and label the point (-2,0)
    \fill[blue] (-2,0) circle (2pt) node[above] {$\alpha$};

    % Draw the line from (-2,0) through (x_i,x_n) to x=1
    \draw[dashed] (-2,0) -- (A) -- (1, 0.9230769231);

    % Forming a sexond triangle for the purpose of the description
    \draw[dashed] (A) -- (0.6,0);

    % Give the intersection point on the line 
    \coordinate (B) at (1, 0.9230769231);

    \fill[red] (B) circle (2pt) node[above right] {$(1, x_i')$};
    
    % Draw the line x = 1 in a different color
    \draw[red] (1,-1.5) -- (1,1.5) node[above right] {$X_n=1$};
\end{tikzpicture}
    \caption{Observe that two right-angled triangles form in this diagram plotting $X_i$ against 
    $X_n$, and derive Equation (\ref{triangleeqn}) from the ratio of their side lengths.} 
    \label{fig:trianglepic} 
\end{figure}

\end{proof}
\end{lemma}

Let $f  : A\to B$ be a function between two sets $A$ and $B$. When $A'\subseteq A$ and $B' \subseteq B$, we denote the \textit{domain restriction} of $f$ to $A'$ by $f|_{A'}$, and the \textit{codomain restriction} of $f$ to $B'$ by $f|^{B'}$ (whenever such a codomain restriction exists).

\begin{proposition}\label{propDomainRange}
    Let $n \in \mathbb{N}$. Take $\alpha \in [-\infty,1)$, and let $P_\alpha \in \mcb(\mathbb{R}^{n+1};\mathbb{R}^n)$ be as in Definition \ref{PalphaDefinition}. Define the sets $S_\alpha$ and $R_\alpha$ as follows.
    
    \begin{enumerate}[(i)]
        \item\label{C1} If $\alpha \in [-1,1)$, set \[S_\alpha = \{ (x_i)_{i=0}^n \in \mathbb{S}_{\mathbb{R}^{n+1}} : x_n > \alpha    \}\;,\] and let $R_\alpha = \mathbb{R}^n$.

        \item\label{C2} If $\alpha \in (-\infty, -1)$, set  \[S_\alpha = \left\{ (x_i)_{i=0}^n  \in \mathbb{S}_{\mathbb{R}^{n+1}} : x_n \geq \frac{1}{\alpha}    \right\}\;,\] and let \[ R_\alpha = \sqrt{\frac{\alpha - 1}{\alpha + 1}}\, B_{\mathbb{R}^n}  \;. \]

        \item\label{C3} If $\alpha = -\infty$, set \[S_\alpha = \{ (x_i)_{i=0}^n  \in \mathbb{S}_{\mathbb{R}^{n+1}} : x_n \geq 0\}\;,\] and let $R_\alpha = B_{\mathbb{R}^n}$.
    \end{enumerate}

    Then $P_\alpha|^{R_\alpha}_{S_\alpha} : S_\alpha \to R_\alpha$ is a bijection.
\begin{proof}
    We begin by showing that $P_\alpha|_{S_\alpha}^{R_\alpha}$ is injective. Notice that there can be at most two points on the surface of the hypersphere $\mathbb{S}_{\mathbb{R}^{n+1}}$ colinear with both the point of projection $(0,0,\dots,0,\alpha)$ and the tangent hyperplane defined by $x_n = 1$. The set $S_\alpha$ is defined specifically to only contain one such point. This tells us that if $x,y \in \mathbb{S}_{\mathbb{R}^{n+1}}$ with $P_\alpha x = P_\alpha y$, then only one of $x$ and $y$ can belong to $S_\alpha$. This proves that $P_\alpha|_{S_\alpha}^{R_\alpha}$ is injective as required.

    Now, we check that $P_\alpha|_{S_\alpha}^{R_\alpha}$ is surjective. Let $y = (y_i)_{i=0}^{n-1} \in R_\alpha$. For $\alpha \in (-\infty,1)$, we notice that a line connecting $(0,\dots,0,\alpha)$ and $(y_0,\dots,y_{n-1},1)$ intersects $S_\alpha$ exactly once, proving the claim. Finally, for $\alpha = - \infty$, taking $x = (y_0,\dots,y_{n-1},x_n)$ where \[x_n = \sqrt{1-\sum_{i=0}^{n-1} y_i^2}\;,\] we see that $x \in S_\alpha$ with $P_\alpha(x)=y$.
\end{proof}
\end{proposition}

As alluded to in the above proof, the subsets $S_\alpha$ of $\mathbb{S}_{\mathbb{R}^{n+1}}$ are specifically defined in order to make $P_\alpha|_{S_\alpha}$ injective. The intuition behind their definitions is that a line which intersects the point of projection $(\alpha,0,\dots,0)$ and the hypersphere $\mathbb{S}_{\mathbb{R}^{n+1}}$, can intersect $\mathbb{S}_{\mathbb{R}^{n+1}}$ at most twice. In cases where there are two such intersection points, the set $S_\alpha$ is chosen to include only one of these two points - the one closest to the plane of projection defined by $x_n = 1$.

\begin{figure}[!h]
    \centering
    \begin{tikzpicture}[scale=1]
    % Draw the axes
    \draw[->] (-3,0) -- (3,0) node[right] {$X_1$};
    \draw[->] (0,-3) -- (0,3) node[above] {$X_0$};

    % Draw and label the point (-2,0)
    \fill[blue, thick] (-1.666667,0) circle (2pt) node[above] {$\alpha$};

    \draw[dashed]  (-1.666667,0) -- (-0.6,0.8) -- (1,2);

    \draw[dashed]  (-1.666667,0) -- (-0.6,-0.8) -- (1,-2);

    %Draw the arc to represent S_alpha
    \draw [ao(english),thick,domain=-126.87:126.87] plot ({cos(\x)}, {sin(\x)}) node[below right]{$S_\alpha$};

    % Draw the line x = 1 with one colour for R_alpha and one for outside of R_alpha
    \draw[red] (1,2) -- (1,3) node[above right] {$X_1=1$};
    \draw[blue] (1,-2) -- (1,2) node[below right] {$R_\alpha$};
    \draw[red] (1,-3) -- (1,-2);
\end{tikzpicture}
    \caption{In the case $n=1$ and $\alpha < -1$, the blue line segment is the set $R_\alpha$, embedded into two dimensional space as tangent to the unit circle. The set $S_\alpha$ is a subset of the unit circle and is depicted in green. The dashed lines are tangents to the unit circle crossing through the point $(\alpha,0)$.}
    \label{fig:EalphaDomainRange} 
\end{figure}
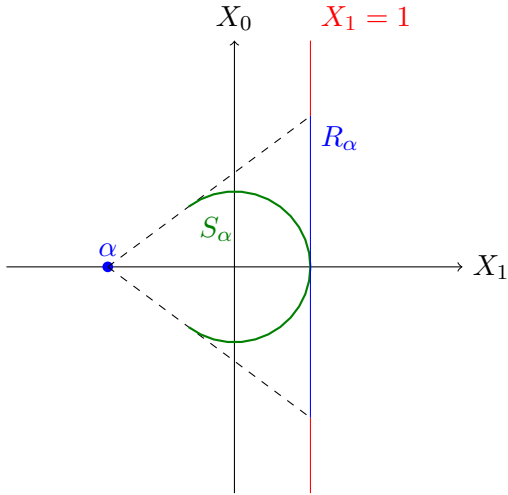

In Figure \ref{fig:EalphaDomainRange}, we display an example for $n=1$, $\alpha < -1$ to help to intuit the sets $R_\alpha$ and $S_\alpha$.

\section{Definition of $E_\alpha$ for $\alpha \in [-\infty,1)$}\label{sectionEmbeddingDefn}
Fix $n \in \mathbb{N}$. Our goal for this section will be to define for every $\alpha \in [-\infty,1)$, a map $E_\alpha : \mathbb{R}^n \to \mathbb{R}^{n+1}$, with range contained in the closed unit sphere $\mathbb{S}_{\mathbb{R}^{n+1}}$ of $\mathbb{R}^{n+1}$. The map $E_\alpha$ will in part be defined to be an inverse of the bijection $P_\alpha|^{R_\alpha}_{S_\alpha}$ as calculated in Proposition \ref{propDomainRange}.

Since the range $P_\alpha$ of  $P_\alpha|^{R_\alpha}_{S_\alpha}$ is bounded whenever $\alpha < -1$, and we require $E_\alpha$ to be defined on the entirety of $\mathbb{R}^n$, we define the action of $E_\alpha$ to map points outside of $R_\alpha$ to points on the boundary of $S_\alpha$. This of course makes the map $E_\alpha$ not injective, however this will not be a problem when utilising $E_\alpha$ as a quantum embedding since in practice, any distinct data points which $E_\alpha$ identifies can always be considered extreme outliers.

That $E_\alpha$ is defined on the entirety of $\mathbb{R}^n$ will be of key importance in Section \ref{sectionEquivariance}, where we wish to consider the interactions between $E_\alpha$ and group representations on $\mathbb{R}^n$ and $\mathbb{R}^{n+1}$.

\begin{definition}\label{defnEalpha}
Let $n \in \mathbb{N}$ and $\alpha \in [-\infty,1)$. Let $P_\alpha$ be as in Definition \ref{PalphaDefinition}, and let $R_\alpha$ and $S_\alpha$ be as in Proposition \ref{propDomainRange}. Define $E_\alpha :\mathbb{R}^n \to \mathbb{R}^{n+1}$ to have action \[ x \mapsto \begin{cases} (P_\alpha|_{S_\alpha}^{R_\alpha})^{-1}(x) & \;\text{ if } x \in R_\alpha\;, \\  (P_\alpha|_{S_\alpha}^{R_\alpha})^{-1}(\beta x) & \; \text{ if } x \in \mathbb{R}^{n}\setminus R_\alpha\;.\end{cases} \]

Here, $\beta = \|x\|^{-1}$ if $\alpha = -\infty$, and \[\beta = \sqrt{\frac{\alpha - 1}{\alpha + 1}} \|x\|^{-1}\] for every $\alpha \in (-\infty, -1)$.
\end{definition}

Proposition \ref{propDomainRange} proves that for every $\alpha \in [-\infty,1)$, $E_\alpha$ is well-defined. Akin to Lemma \ref{concreteDefnPalpha}, for the benefit of future calculations and to reduce the abstractness of our methods, we give our next result, which concretely expresses the final coordinate of the image of an arbitrary point under $E_\alpha$. To this end, for each $n \in \mathbb{N}$, $0\leq i \leq n$, $x \in \mathbb{R}^{n}$ and $\alpha \in [-\infty,1)$, write $E_\alpha(x)_i$ to denote the $i$th coordinate of $E_\alpha(x)$.

\begin{lemma}\label{firstcoordinate}
Fix $n \in \mathbb{N}$, and let $x = (x_i)_{i=0}^{n-1} \in \mathbb{R}^n$. Let $\alpha\in (-\infty,1)$, and let $E_\alpha \in \mcb(\mathbb{R}^{n}; \mathbb{R}^{n+1})$ be as in Definition \ref{defnEalpha}, with $R_\alpha$ and $S_\alpha$ as in Proposition \ref{propDomainRange}. 

If $x \in R_\alpha$, let $s=\|x\|$, and if $x \notin R_\alpha$, let $s = \sqrt{\frac{\alpha-1}{\alpha+1}}$. Then \[E_\alpha (x)_n = \frac{-B + \sqrt{B^2 - 4AC}}{2A}\;,\] where \begin{equation*}\begin{gathered}
    A = s^2 + (1-\alpha)^2\;, \quad B = -2\alpha s\;, \\ C = s^2\alpha^2 - (1-\alpha)^2\;.\end{gathered}
\end{equation*}  
Moreover, if $\alpha = -\infty$ and $x \in R_\alpha$, then \[E_\alpha(x)_n = \sqrt{1-\|x\|^2}\;,\] and if $x \notin R_\alpha$, we have that \[E_\alpha(x)_n = 1\;.\]

In particular, $E_\alpha (x)_n$ is a function of only $\|x\|$ and $\alpha$.

\begin{proof}
    Suppose first that $\alpha \in (-\infty,1)$ and $x\in R_\alpha$. Embed $x$ in the natural way into the hyperplane of $\mathbb{R}^{n+1}$ defined by equation $x_n = 1$, to obtain the vector $x' = (x_0,\dots,x_{n-1},1)$. Then $E_\alpha(x)$ is the point in $S_\alpha$ colinear with both $x'$ and $(0,\dots,0,\alpha)$.

   From Lemma \ref{concreteDefnPalpha}, we see that \[E_\alpha(x)_{i} = \frac{(E_\alpha(x)_n - \alpha)}{(1-\alpha)}x_i \] for every $i \in \{0,\dots,n-1\}$. Now using the fact that $E_\alpha(x)$ has norm $1$, we deduce that \begin{align*}
       E_\alpha(x)_n^2 &= 1 - \sum_{i=0}^{n-1} E_\alpha(x)_i^2 \\ & = 1 -\sum^{n-1}_{i=0}\frac{(E_\alpha(x)_n - \alpha)^2}{(1-\alpha)^2} x_i^2 \\ & = 1- \frac{(E_\alpha(x)_n - \alpha)^2}{(1-\alpha)^2} \|x\|^2\;.
   \end{align*}
   The result follows from using the quadratic formula to find $E_\alpha(x)_n$ from the above equation, selecting the unique solution which belongs to the set $S_\alpha$.

   Similarly, for $\alpha = -\infty$ and $x \in R_\alpha$, we get that \[E_\alpha(x)_n^2 = 1- \| x\|^2\] because the norm of $E_\alpha(x)$ must be equal to $1$. From this, the conclusion follows easily.

   The cases for $x \notin R_\alpha$ follow from applying the same methods to a version of $x$ scaled by $\beta$ to belong to the boundary of $R_\alpha$. Thus $E_\alpha (x)_n$ is a function of only $\|x\|$ and $\alpha$ as claimed. \end{proof}
\end{lemma}

We finish this section with a brief explanation of how the functions $E_\alpha$ for $\alpha \in [-\infty,1)$ can be used as classical-to-quantum data embeddings. The state space of a quantum computer of $n$ qubits is the complex Hilbert space $\mch$ of dimension $2^n$. The set of possible quantum states on said quantum computer is given by $\mathbb{S}_\mch$. Clearly, $\mathbb{R}^{2^n}$ is a subspace of $\mch$, and $\mathbb{S}_{\mathbb{R}^{2^n}}$ is a subset of $\mathbb{S}_\mch$. We henceforth consider $E_\alpha$ as a function with range $\mch$ by identifying $E_\alpha$ with the composition of $E_\alpha$ and the canonical embedding $\mathbb{R}^{2^n}$ to $\mathbb{C}^{2^n}$.

Expressed in standard Dirac notation, for each $n \in \mathbb{N}$, and each $\alpha \in [-\infty,1)$, we map the vector $x \in \mathbb{R}^{2^n-1}$ to the quantum state $| E_\alpha(x)\rangle$ defined as \begin{equation}\label{DiracEalpha} | E_\alpha(x)\rangle := \sum_{i=0}^{2^n-1} E_\alpha(x)_i |i\rangle\;. \end{equation}

\section{Group Equivariance}\label{sectionEquivariance} 

The authors would like to stress that the representation theoretic fundamentals at the beginning of this section are not new, and can be found in e.g., \cite{meyer, larocca}. However, we believe that it is important for a complete understanding to give each result formally alongside a mathematically rigorous proof. 

Further examples of equivariant embeddings in QML can be found in \cite{meyer, nguyen, west}.

\begin{definition}
    Let $G$ be a group, and let $V$ and $W$ be vector spaces. Let $\psi : G \to \mcb (V)$ and $\phi : G \to \mcb (W)$ be representations of $G$ on $V$ and $W$ respectively. 

    A function $f : V \to W$ is said to be \textit{equivariant} with respect to $G$ (with representations $\psi$ and $\phi$) if for all $g \in G$, we have that \begin{equation}\label{EquivariantDefn}
        f(\psi(g) v) = \phi(g) f(v) \,,\qquad \forall v \in V\;.
    \end{equation}
\end{definition}

An important feature of equivariant functions is that the property of equivariance is preserved under composition. The following proposition states this clearly. For brevity, during this section we may omit the precise description of which group representations some functions are equivariant under. This is often clear from context and should not cause confusion.

\begin{proposition}\label{propositionEquivarianceComposition}
    Let \( G \) be a group, and let \( V, W, \) and \( X \) be vector spaces. Let \( \psi : G \to \mcb (V) \), \( \phi : G \to \mcb (W) \), and \( \xi : G \to \mcb (X) \) be representations of \( G \) on \( V, W, \) and \( X \) respectively. Let \( f : V \to W \) and \( h : W \to X \) be equivariant with respect to \( G \). The composition \(  h \circ f : V \to X \) is equivariant with respect to \( G \). 

    It follows that any composition of finitely many functions that are equivariant with respect to $G$, is an equivariant function with respect to $G$.

\begin{proof}
   Since \( f \) and \( h \) are equivariant with respect to \( G \), we have for all \( v \in V \), \(w \in W\), and \( g \in G \), $  f(\psi(g)v) = \phi(g)f(v)$, and 
$h(\phi(g)w) = \xi(g)h(w)$
So, we have \begin{align*}
    (h \circ f)(\psi(g)v) = h(\phi(g) f(v)) & \\ =  \xi(g)h(f(v)) = & \xi(g)(h \circ f)(v)\;.
\end{align*}
This proves that \( h \circ f \) is equivariant with respect to \( G \) as required. 

The second claim pertaining to composing arbitrarily many equivariant functions follows from a simple induction argument.
\end{proof}
\end{proposition}

Now, consider the fact that gates in a quantum circuit are simply manifestations of unitary operators on Hilbert spaces, and expectations of measurements of quantum systems are themselves simply functions from the set of possible quantum states to the real line. If a chosen embedding mapping classical data to a quantum state is equivariant with respect to a given group, and so is every gate and measurement enacted on that state in a quantum circuit, the entire quantum circuit will then be equivariant with respect to said group. In such a circuit, inputs related by some symmetry group will always yield outputs related by the same symmetry (modulo the effects of noise). The upshot of this for QML is that information about symmetry in the data no longer needs to be learned by an equivariant variational quantum circuit - this information about the data is incorporated to the structure of the model even before training. In certain cases, this can lead to dramatic increases in training accuracy and training speed.

Proposition \ref{twirling formula} provides us with a simple method of determining whether a gate is equivariant. For its proof, we require the following basic group theoretic lemma.

\begin{lemma}\label{left-multiplication-bijective}
Let $G$ be a group, and let $h \in G$. The function $L_h : G \to G$ defined by $L_h(g) = hg$ for every $g \in G$, is a bijection. 
%\vspace{-3em}
\begin{proof} Taking $g_1,g_2 \in G$ if $L_h(g_1) = L_h(g_2)$, then \begin{align*}
        g_1 =  h^{-1}h g_1  &= h^{-1}L_h(g_1) \\ &= h^{-1}L_h(g_2) = h^{-1}hg_2 = g_2\;,  
    \end{align*}  so $L_h$ is injective. Further, for all $g \in G$, we have that $g = L_h(h^{-1} g)$, so $L_h$ is surjective.
    \end{proof}
\end{lemma}

We hereon restrict our scope to finite groups. A version of the following proposition for Lie groups exists and can be found in e.g., \cite{meyer}, however this is outside the scope of this paper.

\begin{proposition}\label{twirling formula}
Let $G$ be a finite group, let $V$ be a Hilbert space, and let $\phi : G \to \mcb (V)$ be a unitary representation of $G$ on $V$. The function $\mathcal{T}_\phi : \mcb (V) \to \mcb (V)$ defined by \begin{equation}\label{twirling equation} \mathcal{T}_\phi( U ) = \frac{1}{|G|}\sum_{g \in G} \phi(g) U \phi(g)^*\,,\qquad \forall U \in \mcb (V)  \end{equation} is a projection of $\mcb (V)$ onto the space of operators equivariant with respect to $G$.

\begin{proof}
Let $U \in \mcb (V)$, $v \in V$, and $h \in G$. Then
\begin{align*}
 \mathcal{T}_\phi( U ) \phi(h)v &= \frac{1}{|G|}\sum_{g \in G} \phi(g) U \phi(g)^*\phi(h)v \\
                                &= \frac{ \phi(h)}{|G|} \sum_{g \in G}\phi(h)^* \phi(g) U \phi(g)^*\phi(h)v \\
                                &= \frac{ \phi(h)}{|G|} \sum_{g \in G}\phi(h^{-1} g) U \phi(g^{-1}h)v \\ 
                                & = \frac{ \phi(h)}{|G|} \sum_{x \in G}\phi(x) U \phi(x)^* v \\
                                &= \phi(h) \mathcal{T}_\phi (U) v\;,
\end{align*}
where the penultimate equality holds because Lemma \ref{left-multiplication-bijective} allows us to replace the terms of the summation as shown. Thus $\mathcal{T}_\phi(U)$ is equivariant with respect to $G$. Next, we show that $\mathcal{T}_\phi$ is idempotent. Indeed, \begin{multline*}
    \mathcal{T}_\phi(\mathcal{T}_\phi(U)) \\ = \frac{1}{|G|}\sum_{g \in G} \phi(g)\left(\frac{1}{|G|}\sum_{h \in G} \phi(h) U \phi(h)^*\right)\phi(g)^*\;.
\end{multline*}

Expanding and rearranging, we get
\[
\mathcal{T}_\phi(\mathcal{T}_\phi(U))= \frac{1}{|G|^2}\sum_{g \in G}\sum_{h \in G} \phi(gh) U \phi(gh)^*\;.
\]
Again, two further applications of Lemma \ref{left-multiplication-bijective} allow us to rewrite the the sum above, from which we obtain
\[
\mathcal{T}_\phi(\mathcal{T}_\phi(U)) = \frac{1}{|G|}\sum_{x \in G} \phi(x) U \phi(x)^* = \mathcal{T}_\phi(U)\;.
\]

Further simple direct calculations show that $\mathcal{T}_\phi$ is linear, which completes the proof.
\end{proof}
\end{proposition}

Given a representation $\phi$ of a finite group on a Hilbert space $\mch$, the idempotence of $\mathcal{T}_\phi$ allows us to easily identify the unitary operators on $\mch$ that are equivariant. These are exactly those unitaries $U\in \mcb(\mch)$ satisfying $\mathcal{T}_\phi(U) = U$.

Now armed with the knowledge of how to find equivariant gates for a given group representation, and how equivariant circuits can be constructed through compositions of equivariant gates, measurements, and embeddings, we conclude this section with the final ingredient required for us to utilise our embeddings $E_\alpha$ for equivariant QML - the proof that they are indeed equivariant embeddings.

\begin{theorem}\label{EalphaIsEquivariant}
Let $G$ be a group, let $n \in \mathbb{N}$ and let $\rho$ be any unitary representation of $G$ on $\mathbb{R}^n$. Let $ \iota : G \to \mcb(\mathbb{R})$ be the trivial representation of $G$ on $\mathbb{R}$.

The direct sum $\rho \oplus \iota$ is a unitary representation of $G$ on $\mathbb{R}^{n+1}$, and for any $\alpha \in [-\infty,1)$, $E_\alpha$ is equivariant with respect to $G$ with representations $\rho$ and $\rho \oplus \iota$.
\begin{proof}

The fact that $\rho \oplus \iota$ is a unitary representation of $G$ on $\mathbb{R}^{n+1}$ follows simply from the fact that $\iota$ and $\rho$ are unitary representations of $G$ on $\mathbb{R}$ and $\mathbb{R}^n$ respectively.

Using Lemma \ref{firstcoordinate}, we may take a function \[f : [0,\infty)\times [-\infty,1) \to \mathbb{R}\] for which \[E_\alpha(x)_n = f(\|x\| , \alpha)\] for every $x \in \mathbb{R}^n$ and every $\alpha \in [-\infty,1)$. 

To prove the claim about equivariance, fix $\alpha \in [-\infty, 1)$, set $g \in G$ and let $x \in \mathbb{R}^n$. Because $\rho$ is a unitary representation, we have that \[\|\rho(g) x\| = \|x\|\;.\] Therefore,
 \begin{align}
    (\rho \oplus \iota)&(g) E_\alpha(x)\notag\\& =  \rho(g)(E_\alpha(x)_i)_{i=0}^{n-1} \oplus \iota(g) E_\alpha(x)_n  \notag\\
    &=  \rho(g)(E_\alpha(x)_i)_{i=0}^{n-1}\oplus f(\|x\|,\alpha) \notag\\
    &=\rho(g)(E_\alpha(x)_i)_{i=0}^{n-1} \oplus f(\|\rho(g)(x)\|,\alpha)   \notag \\
    &= \rho(g)(E_\alpha(x)_i)_{i=0}^{n-1} \oplus E_\alpha(\rho(g)(x))_n \;.\label{eqnA}
\end{align}

Consider the definition of the set $R_\alpha$ as in Proposition \ref{propDomainRange}. No matter the value of $\alpha$, $R_\alpha$ is either a hypersphere centred at the origin, or it is the entire space $\mathbb{R}^n$. Since $\rho(g)$ is unitary, it is an isometry, and it follows that $x \in R_\alpha $ if and only if $\rho(g)x \in R_\alpha$. Thus, because $\rho(g)$ is linear, we can hereon restrict ourselves to the case for $x \in R_\alpha$ (else, replace $x$ with $\beta x$ before proceeding, where $\beta $ is as in Definition \ref{defnEalpha}).

Lemma \ref{concreteDefnPalpha} then tells us that
\begin{align}
 \rho(g)((&E_\alpha(x)_i)_{i=0}^{n-1})\notag \\ &= \rho(g) \left(\left(\frac{f(\|x\|,\alpha)-\alpha}{1-\alpha}\right)(x)\right) \notag \\
 &= \left(\frac{f(\|\rho(g)(x)\|,\alpha)-\alpha}{1-\alpha}\right) \rho(g)(x)\notag \\ &=E_\alpha(\rho(g)(x))_{i=0}^{n-1}\;. \label{eqnB}
\end{align}

Combining Equations (\ref{eqnA}) and (\ref{eqnB}) now gives 
\begin{align*}
 (\rho \oplus & \iota)(g) E_\alpha(x) \\&= E_\alpha(\rho(g)(x))_{i=0}^{n-1} \oplus  E_\alpha(\rho(g)(x))_n \\
 &= E_\alpha(\rho(g)(x))\;,\end{align*} completing the proof.
\end{proof}
\end{theorem}

To intuit the above theorem, think of unitary operators on Hilbert spaces as generalised rotations about the origin and reflections about some hyperplane through the origin. Then observe that to rotate or reflect a vector in $\mathbb{R}^n$ and then embed it to the unit hypersphere of $\mathbb{R}^{n+1}$ using $E_\alpha$, is to perform the same action as to embed that vector to the unit hypersphere using $E_\alpha$ and then rotate or reflect it in the same way.

\begin{remark}
    Theorem \ref{EalphaIsEquivariant} considers for each $\alpha \in [-\infty,1)$, $E_\alpha$ as a function from $\mathbb{R}^n$ to $\mathbb{R}^{n+1}$. On the other hand, the final paragraphs of Section \ref{sectionEmbeddingDefn} detail that when using $E_\alpha$ in practice as a quantum embedding, we consider it to have codomain $\mathbb{C}^{n+1}$ by implicitly composing it with the natural embedding $\mathbb{R}^{n+1} \to \mathbb{C}^{n+1}$. Using Proposition \ref{propositionEquivarianceComposition}, we see that the equivariance of $E_\alpha$ is not disturbed by this process, because it is easy to see that the natural embedding $\mathbb{R}^{n+1} \to \mathbb{C}^{n+1}$ is itself equivariant.
\end{remark}

\section{Experiments}\label{sectionExperiments}
\subsection{Setup}
For $\alpha = 0$, $\alpha = -1$, $\alpha = -1-\frac{\sqrt{2}}{2}$, and $\alpha = -\infty$, we term the maps $E_\alpha$ from Definition \ref{defnEalpha} to be the \textit{reverse gnomonic}, \textit{reverse stereographic}, \textit{reverse twilight}, and \textit{reverse orthographic} embeddings respectively. The goal of this section is to display a series of comparisons between these embeddings.

We first do this by investigating their like-for-like performance on a basic machine learning classification task using a typical quantum neural network (QNN). The data for the classification task is chosen with labels that are invariant to a $\mathbb{Z}_2$ symmetry. Leveraging this symmetry, we also investigate the capacity of the four chosen embeddings as equivariant embeddings by performing the same classification task with a second QNN, of equal size, which is equivariant with respect to the symmetry group. A comparison of each embedding with an equivariant and non-equivariant setup can then also be made. We also include standard amplitude embedding in our comparisons.

\begin{figure}[!h]
\centering
\includegraphics[width=0.4\textwidth]{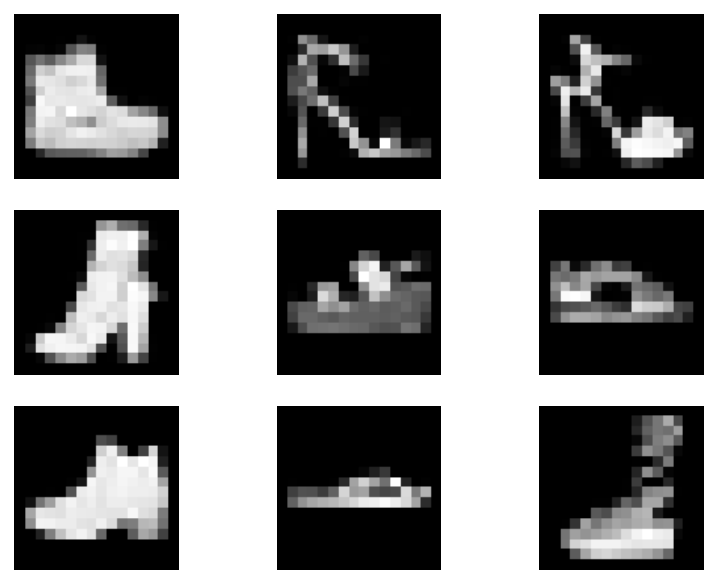}
\caption{\small{Footwear data from the fashion MNIST database \cite{xiao2017fashion}. The task for the quantum neural network is to distinguish sandals (open footwear) vs boots (closed footwear). The images have been pixelated to contain $16\times 16 = 2^8$ pixels.}}\label{fig:data} 
\end{figure}

The classification task with which to test our embeddings is to distinguish images of boots from images of sandals, with the footwear randomly directed to point either left or right. The reason behind this random flipping is that the equivariant QNN may have an advantage over the non-equivariant QNN when we specifically construct it to respect the symmetry of horizontally flipping the data. A sample of the data used is shown in Figure \ref{fig:data}.

Because sandals are open footwear and boots are not, we can expect the greyscale images of boots to be more predominantly white in colour when compared to the images of sandals. As such, the data vector for a sandal will typically have a smaller euclidean norm than that of a boot. This norm value features prominently when using a reverse map embedding, but is lost information under standard amplitude encoding. Therefore, the use of reverse map embeddings should help the QNNs to distinguish between the boots and sandals.

\begin{figure}[!h]
    \centering
    \subfigure[Non-Equivariant]{
        \includegraphics[width=0.2\textwidth]{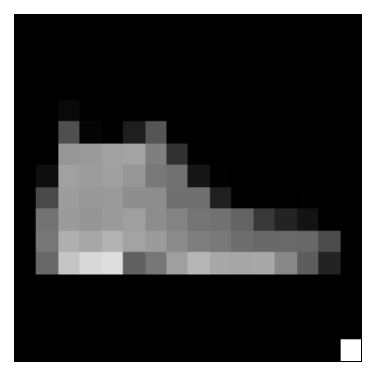}
    }
    \hspace{0.01\textwidth} % Adjust the space between the subfigures
    \subfigure[Equivariant]{
        \includegraphics[width=0.2\textwidth]{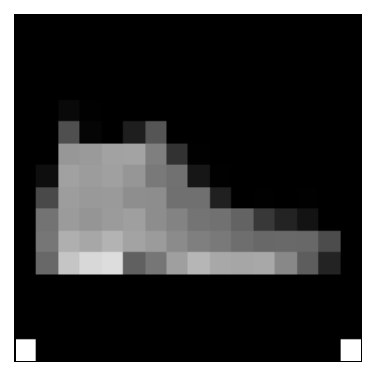}
    }
    \caption{\small{Example data for the Reverse Twilight embedding illustrating - (a) That the final pixel was removed to make way for the additional coordinate from $E_\alpha$. (b) That this pixel is duplicated for the dataset in the equivariant case, so as to not serve as an indicator for which way the shoe is pointed. }}
    \label{fig:data_pix} 
\end{figure}

Given that the codomains of the embeddings $E_\alpha$ have an additional dimension to their domain, and the state space of the quantum computer is always of dimension $2^n$ for some $n \in \mathbb{N}$, it will be beneficial for us to work with data of dimension $2^n - 1$. To this end, we resized the images in the dataset to have $16 \times 16 = 2^8 $ pixels, and before applying the embeddings $E_\alpha$, we also removed their final pixels. This pixel is always a black pixel with value $0$, so its removal does not destroy valuable information. In the equivariant case, this pixel is duplicated in its symmetric position so as not to serve as an indicator for which way the shoe is pointed, and the vector is then renormalised. An illustration of how these pixels manifest is shown in Figure \ref{fig:data_pix}.

Take $\alpha \in [-\infty,-1)$. The radius of $R_\alpha$ as in Proposition \ref{propDomainRange} is $\sqrt{\frac{\alpha - 1}{\alpha + 1}}$ if $\alpha \neq -\infty$, and is $1$ for $\alpha = -\infty$. A key ingredient of Definition \ref{defnEalpha} is that data not contained in $R_\alpha$ is first scaled by some constant $\beta$ before the partial inverse of $P_\alpha$ is applied to it. It is desirable for this to exclusively happen in the case of extreme outliers, so for experiments, before applying $E_\alpha$ to the image data, we scale all the data down by dividing by some scalar $M$. The maximal euclidean norm of the data in our experiment is around $2500$. Thus we set $M = 800$ before applying the reverse twilight embedding, and $M=2500$ before applying the reverse orthographic embedding. For the cases of the reverse gnomonic and reverse stereographic embedding, $\alpha \geq -1$, so $R_\alpha$ is the entire vector space, and no scaling is required. In these cases, results are displayed for the case where no scaling is applied ($M=1$) alongside the case for a somewhat arbitrary scaling factor of $M=2000$, with the goal of demonstrating the profound effect that this scaling factor can have on training. Details on the QNN and training procedure can be found in Appendix \ref{app:QNNdeets}.

\subsection{Non-Equivariant Model}

\begin{figure}[h]
    \centerline{
    \scalebox{0.55}{
        \Qcircuit @C=0.5em @R=0.5em {
            & \gate{R_y(\theta_1)} & \gate{R_z(\theta_2)} & \qw & \qw & \qw & \qw & \qw & \qw & \ctrl{1} & \gate{R_y(\theta_{17})} & \gate{R_z(\theta_{18})} & \qw \\
            & \gate{R_y(\theta_3)} & \gate{R_z(\theta_4)} & \qw & \qw & \qw & \qw & \qw & \ctrl{1} & \targ & \gate{R_y(\theta_{19})} & \gate{R_z(\theta_{20})} & \qw \\
            & \gate{R_y(\theta_5)} & \gate{R_z(\theta_6)} & \qw & \qw & \qw & \qw & \ctrl{1} & \targ & \qw & \gate{R_y(\theta_{21})} & \gate{R_z(\theta_{22})} & \qw \\
            & \gate{R_y(\theta_7)} & \gate{R_z(\theta_8)} & \qw & \qw & \qw & \ctrl{1} & \targ & \qw & \qw & \gate{R_y(\theta_{23})} & \gate{R_z(\theta_{24})} & \qw \\
            & \gate{R_y(\theta_9)} & \gate{R_z(\theta_{10})} & \qw & \qw & \ctrl{1} & \targ & \qw & \qw & \qw & \gate{R_y(\theta_{25})} & \gate{R_z(\theta_{26})} & \qw \\
            & \gate{R_y(\theta_{11})} & \gate{R_z(\theta_{12})} & \qw & \ctrl{1} & \targ & \qw & \qw & \qw & \qw & \gate{R_y(\theta_{27})} & \gate{R_z(\theta_{28})} & \qw \\
            & \gate{R_y(\theta_{13})} & \gate{R_z(\theta_{14})} & \ctrl{1} & \targ & \qw & \qw & \qw & \qw & \qw & \gate{R_y(\theta_{29})} & \gate{R_z(\theta_{30})} & \qw \\
            & \gate{R_y(\theta_{15})} & \gate{R_z(\theta_{16})} & \targ & \qw & \qw & \qw & \qw & \qw & \qw & \gate{R_y(\theta_{31})} & \gate{R_z(\theta_{32})} & \qw \\
        }
    }
    }
    \caption{One repetition of the hardware-efficient ansatz for eight qubits.}
    \label{fig:hardware_efficient_ansatz}
\end{figure}
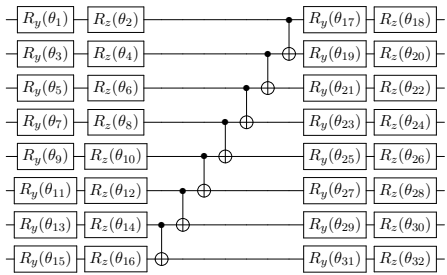

The non-equivariant QNN model architecture consists of our selected embedding followed by one repetition of the hardware-efficient ansatz on $8$ qubits, shown in Figure \ref{fig:hardware_efficient_ansatz}, followed by Pauli-$Z$ measurements on all qubits. 

\subsection{Equivariant Model}

Consider the coordinates $(x,y)$ of a pixel in a given image in our dataset, so that $0\leq x,y < 16$. Express $x$ and $y$ in binary, so that they each have four digits. When we apply $E_\alpha$, viewing through the lens of Equation (\ref{DiracEalpha}), the value of pixel $(x,y)$ corresponds to the amplitude of the orthonormal basis vector $|x\rangle \otimes |y\rangle$. The symmetrically related (under horizontal flipping) coordinate is $(x',y)$, where $x'$ has the bit-flipped binary expression to $x$. To map $|x\rangle \otimes |y\rangle$ to $|x'\rangle \otimes |y\rangle$, or vice versa, we thus apply the tensor product $X_0 \otimes X_1 \otimes X_2 \otimes X_3$ of Pauli-$X$ gates to the first four qubits.

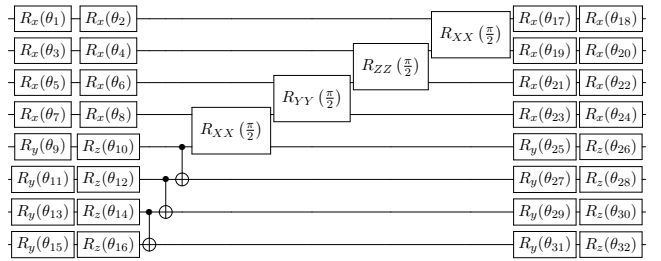
\begin{figure}[h]
    \centerline{
    \scalebox{0.55}{
        \Qcircuit @C=0.2em @R=0.4em {
            & \gate{R_x(\theta_1)} & \gate{R_x(\theta_2)} & \qw & \qw & \qw & \qw & \qw & \qw & \multigate{1}{R_{XX}\left(\frac{\pi}{2}\right)} & \gate{R_x(\theta_{17})} & \gate{R_x(\theta_{18})} & \qw \\
            & \gate{R_x(\theta_3)} & \gate{R_x(\theta_4)} & \qw & \qw & \qw & \qw & \qw & \multigate{1}{R_{ZZ}\left(\frac{\pi}{2}\right)} & \ghost{R_{XX}\left(\frac{\pi}{2}\right)} & \gate{R_x(\theta_{19})} & \gate{R_x(\theta_{20})} & \qw \\
            & \gate{R_x(\theta_5)} & \gate{R_x(\theta_6)} & \qw & \qw & \qw & \qw & \multigate{1}{R_{YY}\left(\frac{\pi}{2}\right)} & \ghost{R_{ZZ}\left(\frac{\pi}{2}\right)} & \qw & \gate{R_x(\theta_{21})} & \gate{R_x(\theta_{22})} & \qw \\
            & \gate{R_x(\theta_7)} & \gate{R_x(\theta_8)} & \qw & \qw & \qw & \multigate{1}{R_{XX}\left(\frac{\pi}{2}\right)} & \ghost{R_{YY}\left(\frac{\pi}{2}\right)} & \qw & \qw & \gate{R_x(\theta_{23})} & \gate{R_x(\theta_{24})} & \qw \\
            & \gate{R_y(\theta_9)} & \gate{R_z(\theta_{10})} & \qw & \qw & \ctrl{1} & \ghost{R_{XX}\left(\frac{\pi}{2}\right)} & \qw & \qw & \qw & \gate{R_y(\theta_{25})} & \gate{R_z(\theta_{26})} & \qw \\
            & \gate{R_y(\theta_{11})} & \gate{R_z(\theta_{12})} & \qw & \ctrl{1} & \targ & \qw & \qw & \qw & \qw & \gate{R_y(\theta_{27})} & \gate{R_z(\theta_{28})} & \qw \\
            & \gate{R_y(\theta_{13})} & \gate{R_z(\theta_{14})} & \ctrl{1} & \targ & \qw & \qw & \qw & \qw & \qw & \gate{R_y(\theta_{29})} & \gate{R_z(\theta_{30})} & \qw \\
            & \gate{R_y(\theta_{15})} & \gate{R_z(\theta_{16})} & \targ & \qw & \qw & \qw & \qw & \qw & \qw & \gate{R_y(\theta_{31})} & \gate{R_z(\theta_{32})} & \qw \\
        }
    }
    }
    \caption{The equivariant ansatz for eight qubits. Compared with the hardware-efficient ansatz, $R_{XX}$ gates replace the CNOT gates, and $R_{X}$ gates replace the $R_{Y}$ and $R_{Z}$ gates, on the first four qubits. Pauli X measurements are made on the first 4 qubits, with Pauli Z measurements on the remaining qubits.}
    \label{fig:equiv_ansatz}
\end{figure}

Denote the horizontal flip operation on the data space by $H$, and the identity operation by $I$. Then $\{ I, H\}$ is a group isomorphic to $\mathbb{Z}_2$. It is easy to see that the function $\phi$ from $\{I,H\}$ to the state space of an $8$ qubit quantum computer which sends $I$ to the identity map and $H$ to $X_0 \otimes X_1 \otimes X_2 \otimes X_3$, is a unitary representation. Theorem \ref{EalphaIsEquivariant} shows that for every $\alpha \in [-\infty,1)$, $E_\alpha$ is equivariant with respect to $\phi$. An application of Proposition \ref{twirling formula} is used to find a set of equivariant gates, and from this, the equivariant QNN architecture is set as shown in Figure \ref{fig:equiv_ansatz}. 

On the first four qubits, $R_{XX}(\frac{\pi}{2})$, $R_{YY}(\frac{\pi}{2})$, and $R_{ZZ}(\frac{\pi}{2})$ gates replace the CNOT gates from the non-equivariant QNN, and $R_{X}$ gates replace all $R_{Y}$ and $R_{Z}$ gates. Pauli-$X$ measurements are performed on the first $4$ qubits, with Pauli-$Z$ measurements on the remaining qubits.

%As an initial test of the equivariant model, we test it against the non-equivariant model using standard amplitude embedding. The results of this test are shown in Table \ref{tab:perform}, where it can be seen that the equivariant QNN significantly improves the classification accuracy. 

\begin{table*}[t]
\centering
\begin{tabularx}{\textwidth}{| >{\raggedright\arraybackslash}m{4cm} 
                              | >{\raggedright\arraybackslash}m{1.5cm} 
                              | >{\raggedright\arraybackslash}m{1cm} 
                              || >{\raggedright\arraybackslash}X 
                              | >{\raggedright\arraybackslash}X |}
\hline
\multicolumn{3}{|c||}{} & \multicolumn{2}{c|}{Test Accuracy (\%)} \\
\hline
\textbf{Embedding} & $\alpha$ & $M$ & \textbf{Non-Equivariant} & \textbf{Equivariant} \\
\hline
Amplitude            & N/A           & N/A        & $80.92 \pm 3.27$ & $92.41 \pm 4.00$ \\
\hline
Reverse Orthographic & $-\infty$     & $2500$     & $83.83 \pm 1.50$ & $89.33 \pm 2.00$ \\
\hline
Reverse Twilight     & $-1- \frac{\sqrt{2}}{2}$ & $800$ & $91.67 \pm 0.00$ & $93.16 \pm 0.50$ \\
\hline
Reverse Stereographic & $-1$         & $1$        & $53.83 \pm 1.50$ & $52.66 \pm 2.00$ \\
\hline
Reverse Stereographic & $-1$         & $2000$     & $89.50 \pm 1.50$ & $94.67 \pm 1.00$ \\
\hline
Reverse Gnomonic     & $0$           & $1$        & $84.50 \pm 1.50$ & $89.17 \pm 2.50$ \\
\hline
Reverse Gnomonic     & $0$           & $2000$     & $92.33 \pm 3.00$ & $89.50 \pm 1.50$ \\
\hline
\end{tabularx}
\caption{The non-equivariant and equivariant QNN prediction accuracy on the test dataset with different embedding mappings $E_\alpha$. The scalar $M$ is the factor that the data is divided by before embedding. This choice is irrelevant in the case of amplitude embedding since for every $M >0$, the amplitude embedded states for any vectors $x$ and $x/M$ are equal. Errors included are the standard deviations from 10 training and testing repeats.}
\label{tab:perform}
\end{table*}

\begin{figure}[h]
\centering
\includegraphics[width=0.48\textwidth]{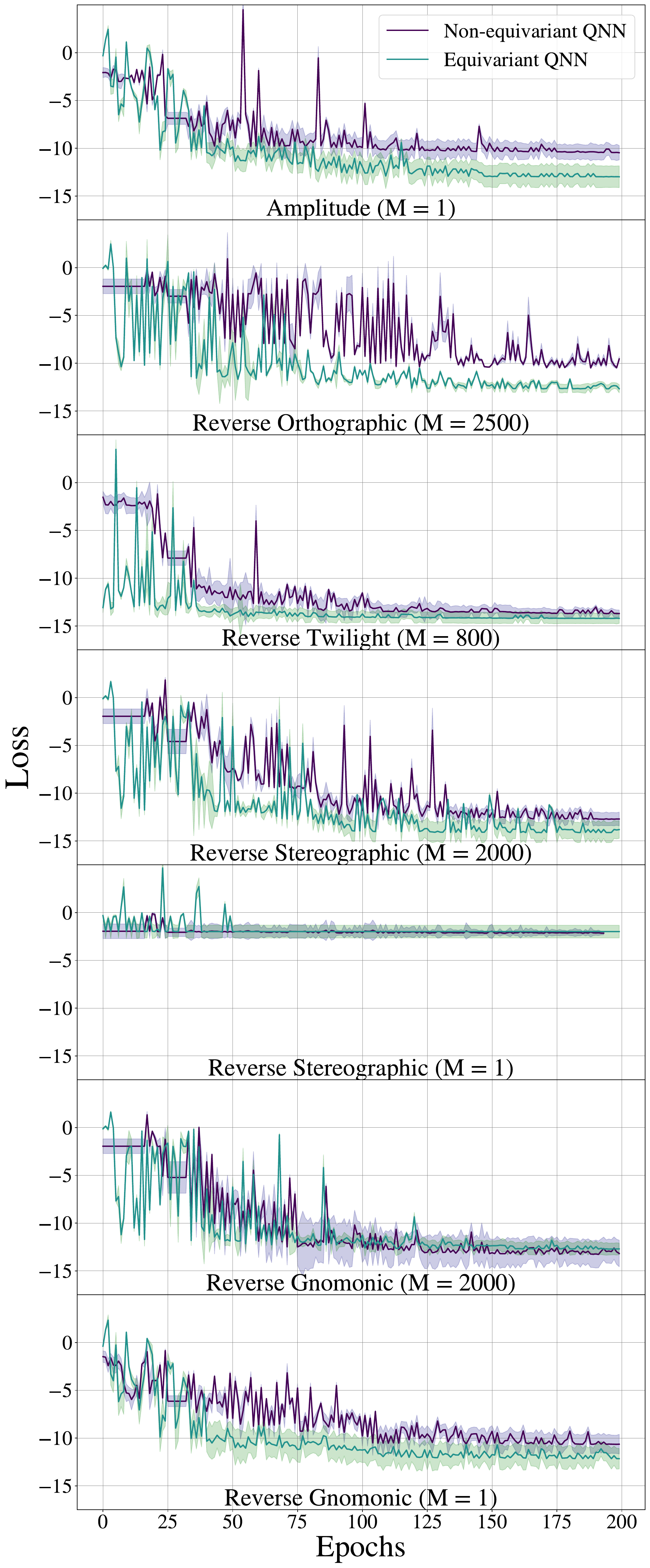}
\caption{\small{The cross-entropy loss over training iterations (epochs) for the non-equivariant and equivariant models. Different embedding mappings $E_\alpha$ are shown with the scaling factor, $M$, used. The shaded areas correspond to the standard deviation errors,
 from $10$ training and testing repeats.
}}\label{plot:final_losses} 
\end{figure}
\twocolumn

\subsection{Results}
The test accuracies for the two circuits across the range of scaling factors and embedding choices are displayed in Table \ref{tab:perform}, and the corresponding loss curve evolutions are shown in Figure \ref{plot:final_losses}. We can see that the equivariant model almost always outperforms the non-equivariant model in terms of accuracy on the test dataset. From the plots in Figure \ref{plot:final_losses}, we also observe that in general, the loss curves for the equivariant QNN reach lower values than their non-equivariant counterparts, and further, they tend to stabilise sooner.

\section{Discussion}

In summary, we defined for every $n \in \mathbb{N}$ and every $\alpha \in [-\infty, 1)$, an embedding $E_\alpha$ from $\mathbb{R}^n$ to the unit hypersphere of $\mathbb{R}^{n+1}$. We showed that $E_\alpha$ is equivariant with respect to any given unitary representation of a group $G$ on $\mathbb{R}^n$ and some unitary representation on $\mathbb{R}^{n+1}$. We then considered $E_\alpha$ as a mapping of classical data into quantum states, comparing the performance of these embeddings on a machine learning classification  task. The capacity of the embeddings $E_\alpha$ as equivariant embeddings for QML was also investigated, where we showed how equivariant quantum circuits can be constructed with $E_\alpha$ as the embedding, displaying that these methods can be employed to improve performance on the classification task.

Considering the results presented in Section \ref{sectionExperiments}, the test accuracy of the the equivariant QNN with amplitude embedding is notably higher than that of the non-equivariant QNN with amplitude embedding. However, with certain reverse map projections and scaling factors, the discrepancy between the equivariant and non-equivariant experimental results seems to be less pronounced, or in some cases is even reversed.

In the future, it would be interesting to investigate the following question: `for a dataset in $\mathbb{R}^n$ on which we wish to perform a QML task using some reverse map projection $E_\alpha$, which value of $\alpha$ should we choose to obtain the best results?' We suspect that this will depend heavily on the nature of the data at hand. Key to this will be to realise why the `boots vs sandals' classification task was so much more successful with some embeddings than with others.

Further, as noise was not present in our simulated experiments, we would like to understand whether there is a relationship between the noise patterns of a quantum computer and the optimal choice of reverse map projection $E_\alpha$ for a given task. We wonder whether certain reverse map projections are more resilient to noise than others.

When moving from a non-equivariant QNN to an equivariant QNN with similar structure, ansatz expressivity may change. In \cite{sim2019expressibility}, expressivity is defined as a circuit's ability to generate (pure) states that are well representative of the Hilbert space. The performance of QNNs, and variational algorithms in general, have been shown to be dependent on the expressivity of the ansatz used \cite{wu2021expressivity,du2022efficient}. As we utilised a different gateset for the equivariant model than for the non-equivariant model, it is difficult to isolate the effects of equivariance without changing the expressivity. We recognise this caveat, attempting to mitigate it by keeping the circuit depth constant, and leave quantifying any expressivity change when moving to an equivariant model for future work.

\section*{Acknowledgements} The authors would like to express their sincere gratitude to Bambord\'{e} Bald\'{e} at Zaiku Group Ltd. and to Phalgun Lolur and Konstantinos Georgopoulos at the UK National Quantum Computing Centre (NQCC) for facilitating this collaboration, and for their guidance and suggestions throughout the project. Their help has been crucial to the successful completion of this work.

This work was supported by the National Quantum Computing centre NQCC200921; and Innovate UK grant 10072286.

\section*{Code availability}
The code used to generate the experimental results presented can be found in the public GitHub repository \cite{equivariant_qnns}.

\bibliographystyle{plain}

\appendix

\section{QNN and Training Details} \label{app:QNNdeets}
To amplitude encode the image vectors into the quantum circuit, Qiskit's `RawFeatureVector' function was used. The QNN was constructed using Qiskit's `NeuralNetworkClassifier' class, and the non-equivariant and equivariant ansatz circuits were given as input to the `EstimatorQNN' function \cite{qiskit2024}. As default, a state vector solver calculated the exact expectation values. The chosen classical optimizer was COBYLA \cite{powell}, along with the cross-entropy loss (objective) function, which is commonly used in machine learning for classification tasks due to the favourable loss landscapes it tends to produce \cite{wang}. The training and testing combined dataset size was $260$ images, with a test/train split of $0.3$. We note that increasing the dataset size could improve training and testing accuracy.

An equivariant QNN with reverse twilight mapping and scaling factor $M = 800$, took $297.96 \pm 79.45$ seconds to train, and $1.085 \pm 0.23$ seconds to test. All other experiments ran for a similar duration. The errors given are from the standard deviation calculated after 10 repeats. Training and testing was performed on a high performance machine with $32$ cores (AMD EPYC 7452 $32$-Core Processor) and $130\unit{Gb}$ of RAM.

\end{document}